\documentclass[a4paper,11pt]{article}
\usepackage{jinstpub} 
\usepackage{lineno}
\usepackage{booktabs}

\title{\boldmath Low dose gamma irradiation study of ATLAS ITk MD8 diodes}

\author[a,1]{M. Mikeštíková,\note{Corresponding author.}}
\author[b]{V. Fadeyev,}
\author[a]{P. Federičová,}
\author[c]{P. Gallus,}
\author[a]{J. Kozáková,}
\author[a]{J. Kroll,}
\author[a]{M. Kůtová,}
\author[a]{J. Kvasnička,}
\author[a]{P. Tůma,}
\author[d]{M. Ullán}
\author[e]{Y. Unno}
\affiliation[a]{Institute of Physics, Czech Academy of Sciences, Na Slovance 2, 18200 Prague, Czech Republic}
\affiliation[b]{Santa Cruz Institute for Particle Physics (SCIPP), University of California, Santa Cruz, CA 95064, USA}
\affiliation[c]{UJP PRAHA a.s., Nad Kamínkou 1345, 156 10 Prague – Zbraslav, Czech Republic}
\affiliation[d]{Instituto de Microelectrónica de Barcelona (IMB-CNM), CSIC,Campus UAB-Bellaterra, 08193 Barcelona, Spain}
\affiliation[e]{Institute of Particle and Nuclear Study, High Energy Accelerator Research Organization (KEK), 1-1 Oho, Tsukuba, Ibaraki 305-0801, Japan}

\emailAdd{Marcela.Mikestikova@cern.ch}

\abstract{Silicon strip detectors developed for the Inner Tracker (ITk) of the ATLAS experiment will operate in a harsh radiation environment of the HL-LHC accelerator. The ITk is thus designed to endure a total fluence of 1.6 $\times$ 10$^{15}$  1-MeV n$_{\mathrm{eq}}/$cm$^2$ and a total ionizing dose (TID) of 66~Mrad in the strip detector region. A radiation-hard $n^+$-in-$p$ technology is implemented in the ITk strip sensors. To achieve the required radiation hardness, extensive irradiation studies were conducted during sensor development, primarily performed up to the maximal expected total fluence and TID to ensure the full functionality of the detector at its end-of-life. These studies included irradiations of sensors with various particle types and energies, including the $^{60}$Co $\gamma$-rays. Our previous results obtained for~$\gamma$-irradiated diodes and strip sensors indicate a linear increase of bulk current with TID, while the~surface current saturates at the lowest TID levels checked (66 Mrad), preventing a determination of the~exact TID for which the observed saturation occurs. 

This work presents the results coming from irradiations by $^{60}$Co $\gamma$-rays to multiple low TIDs, ranging from 0.5 to 100 krad. The detailed study of total, bulk, and surface currents of diodes explores an unknown dependence of surface current on the~TID, annealing, and temperature. Additionally, the effect of the $p$-stop implant between the bias and the guard ring of measured samples is shown. The observations are~relevant for the initial operations of the new ATLAS tracker.}

\keywords{$^{60}$Co gamma irradiation, bulk current, surface current, annealing, temperature dependence}

\begin{document}
\maketitle
\flushbottom



\section{Introduction} \label{sec:intro}
Silicon tracking detectors \cite{bib:STRIPTDR, 1} at the High-Luminosity LHC (HL-LHC) will be subjected to intense radiation, which induces electrically active defects in the silicon.                 
Radiation damage arises primarily through two mechanisms: Non-Ionizing Energy Loss (NIEL) and Total Ionizing Dose (TID).  Non-ionizing effects are referred to as “bulk damage” and are caused by displacement of crystal atoms which therefore leads to the generation of silicon lattice defects, the formation of point and cluster defects, and consequently an increase in deep-level trap states near the middle of the band gap~\cite{pdg2024}. In our study, silicon samples were irradiated with $^{60}$Co gamma rays, where displacement damage is caused by Compton electrons ($ \leq 1~$ MeV) producing only point defects - since the energy is insufficient to create defect clusters.  On a macroscopic scale, NIEL damages in solid state detectors cause an increase of leakage current (increase in noise),  changes in material resistivity, the worsening of the charge collection properties due to trapping mechanism and eventually, the decrease of the carrier’s mobility and their lifetime. On the other hand, ionizing effects affect the insulating layer SiO$_2$ and at the Si-SiO$_{2}$ interface region. These effects are usually referred to as “surface damage” and are responsible for the build-up of trapped charge in the oxide, an increase in bulk oxide traps and interface traps~\cite{Sze}
. In contrast to the silicon bulk, the creation of electron-hole pairs is not fully reversible in insulating materials such as the SiO$_2$ passivation layer, where some electron-hole pairs become separated under an electric field. Due to very different mobility of electrons and holes most of the free electrons leave the SiO$_2$, whereas the holes can drift toward the Si-SiO$_2$ interface and become trapped, leading to an accumulation of positive oxide charges and interface states. The accumulation of positive charge attracts the negative charge from the bulk and the conductive electron channels are created. These effects lead to increased surface leakage currents and deteriorated sensor isolation.  
In earlier measurements on $p$-in-$n$ types of sensors it was observed that the density of oxide charges shows saturation at doses of the order of 100 krad~\cite{bib:renate}. The saturation value is technology-dependent and is larger if the detector is biased during irradiation.



Our previous studies of $\gamma$-irradiated $n^+$-in-$p$ diodes indicate that the bulk current increases linearly with TID, while the surface current saturates~\cite{2, 3}. Unfortunately, the minimum TID in these studies was 66 Mrad, making it impossible to determine at what TID the surface current saturates. Studies performed on $\gamma$-irradiated ATLAS ITk strip main sensors of the ATLAS18 type~\cite{bib:mikestik-vertex} and ITk strip modules using the equivalent sensors~\cite{bib:duden} showed a steep increase of the total current already at low TIDs of 11~krad.  However, the ATLAS18 main sensors do not have a contactable guard ring, thus the surface and bulk current can not be measured separately.

 The presented study is based on precise measurements of the special ITk MD8 diodes irradiated by $^{60}$Co $\gamma$-rays to various low TIDs, performed with separation of the surface and bulk currents. The study also investigates the dependence of these currents on annealing and temperature.


\section{Samples and irradiation} \label{sec:samples}
\begin{figure}[h]
\centering
\raisebox{10mm}{\includegraphics[width=4.8cm]{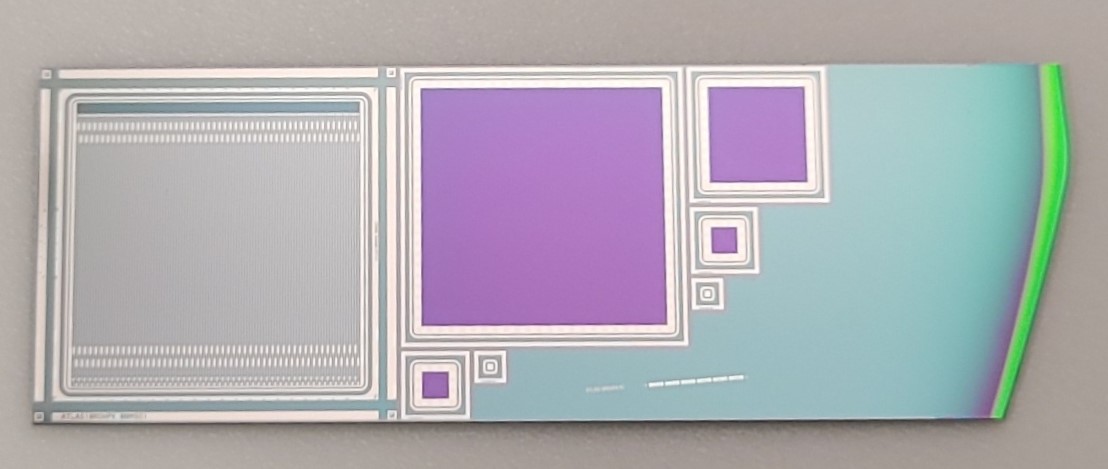}}
\qquad
\includegraphics[width=4.8cm,height=5.5cm, keepaspectratio]{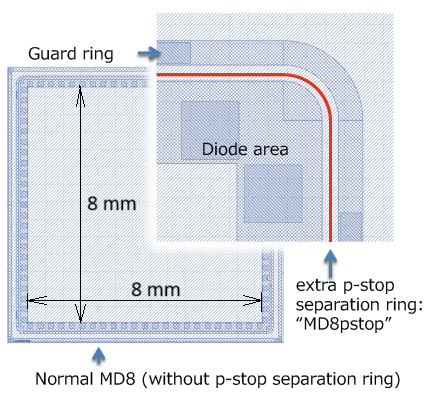}
\caption{{(a) The test structure ``Mini\&MD8'' includes a miniature strip sensor, an MD8 diode (0.8 cm $\times$ 0.8 cm), and several other smaller diodes. (b)~The layout of MD8 and MD8p diodes, which is the same except for the extra p-stop implant between BR and GR ring in MD8p.  \cite{bib:unno}}}\label{fig:vzorky}

\end{figure}

The study was carried out on two types of $n^+$-in-$p$ standard float zone diodes from the ATLAS18 ITk strip production wafers \cite{bib:unno} manufactured by Hamamatsu Photonics K. K. \cite{bib:hamamatsu}: regular MD8 diodes (0.8 cm $\times$ 0.8 cm) without a $p$-stop implant (MD8), and MD8p diodes, which include a $p$-stop implant between the bias ring (BR) and the guard ring (GR). Both diode types have an active area of 0.545 cm$^2$ and a thickness of 320~{\textmu}m.  The additional p-stop implant ring (shown in red colour in Figure~\ref{fig:vzorky}(b) isolates the currents that flow in the diode area from the edge region, especially of surface one, thus providing a better-defined diode volume and reducing edge current interference. The MD8p design allows for more precise studies of radiation-induced damage. In the study, the MD8 diodes were part of a structure called ``Mini\&MD8'', which integrates a miniature strip sensor and an MD8 diode on a single silicon piece \cite{bib:ullan}, see Figure~\ref{fig:vzorky}(a). The MD8p diodes were diced individually.

The samples were irradiated by a $^{60}$Co $\gamma$-source at UJP PRAHA a.s. \cite{bib:ujp} to 14 different TIDs, ranging from 0.5 to 100 krad. During the irradiation, the diodes were placed in a Pb/Al charge particle equilibrium box, as recommended by ESA/SCC \cite{bib:esa}. This setup minimizes dose enhancement by establishing electron equilibrium, ensures a uniform distribution of energy deposition in the irradiated diodes, and reduces the additional dose from low-energy scattered radiation. The samples were irradiated in two irradiation campaigns with the dose rate (in silicon) of 1.60 krad/min (up to 8 krad) and 8.5 krad/min (from 10 to 100 krad), with an estimated uncertainty of less than 5 \%. This in-silicon dose rate was calculated from the in-air dose rate measured by a calibrated ionizing chamber under the same conditions as the tested samples. Cooling with an air fan kept the temperature below  {35}$^\circ$C throughout the irradiation. After the irradiation, the samples were promptly refrigerated at temperatures below  {-20}~$^\circ$C to prevent uncontrolled annealing. 



\section{Experimental methods and results} \label{sec:methods}

To evaluate the effects of radiation damage induced by $^{60}$Co $\gamma$-rays, the \textit{I--V} characteristics of the diodes were measured before and after their irradiation. Great care was taken to properly determine the bulk leakage current flowing exclusively through the active volume of the diode ($I_{\mathrm{BULK}}$) bounded by the GR and the current flowing from GR to the edge that is dominated by the surface current ($I_{\mathrm{SURF}}$), see Figure 1 in \cite{2}.

The tests were carried out using a probe station with controlled temperature {(20$\pm$0.1)}~$^\circ$C and relative humidity ({{<1}\%}) for both unirradiated and irradiated diodes. The irradiated diodes were tested before and after annealing, which was performed for 80 minutes at  {60}~$^\circ$C. 

Furthermore, the MD8p diode irradiated to 25 krad underwent a controlled annealing at  {60}~$^\circ$C for durations of 10, 20, 40, 60, 80, 160, 320, 640, and 1280 minutes. Isochronal annealing was also performed, with 20-minute steps at temperatures increasing from  {60}~$^\circ$C to  {300}~$^\circ$C, using an MD8p diode irradiated to 15 krad. The temperature dependence of the surface current was studied using three MD8p diodes irradiated to TIDs of 10, 50, and 100 krad, over a temperature range from  {-50}~$^\circ$C to  {+20}~$^\circ$C.

\subsection{Bulk, surface, and total currents measured before and after irradiation} \label{sec:currents}

\begin{figure}[h]
\centering
\includegraphics[width=7.3cm]{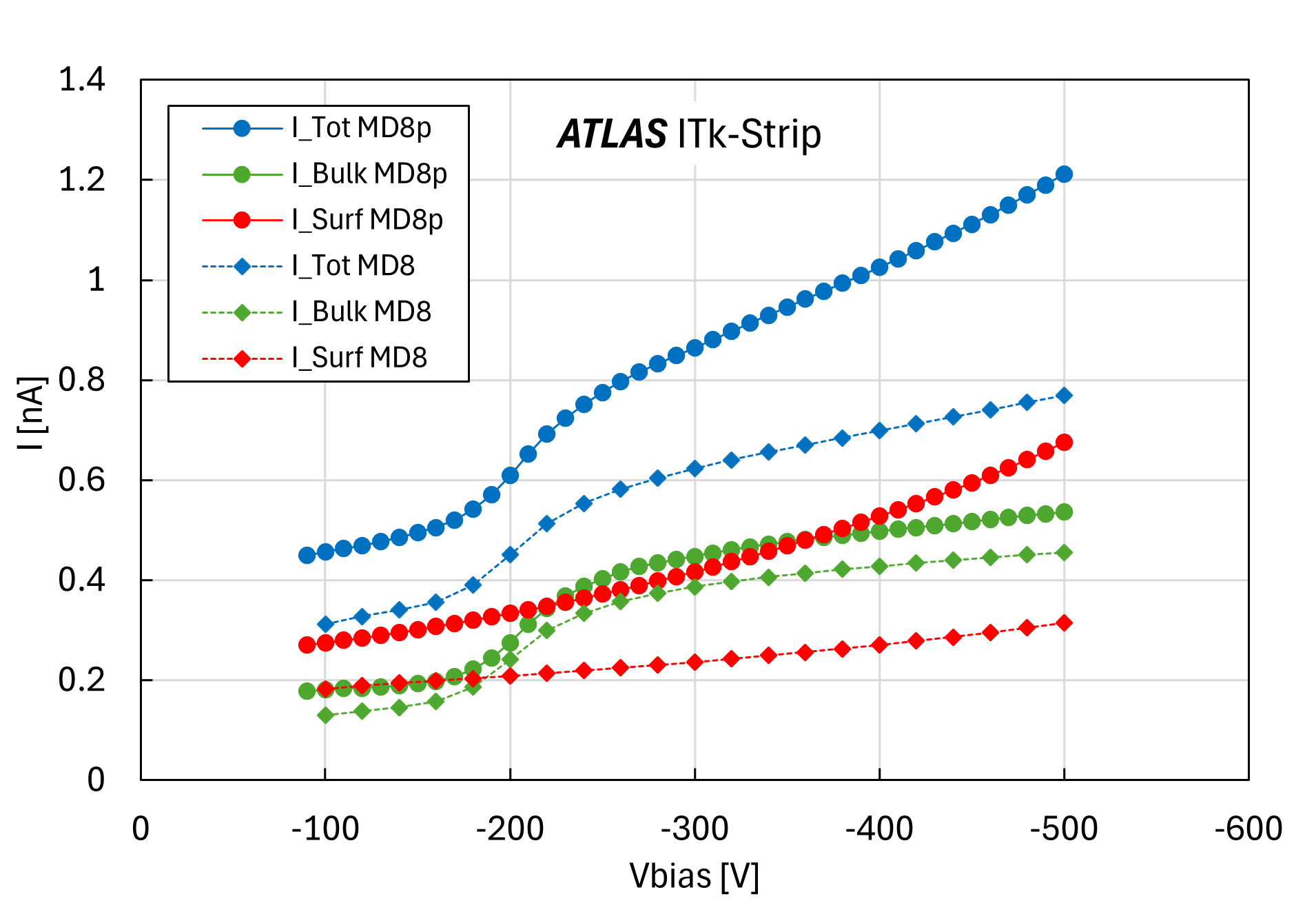} 
\includegraphics[width=7.4cm]{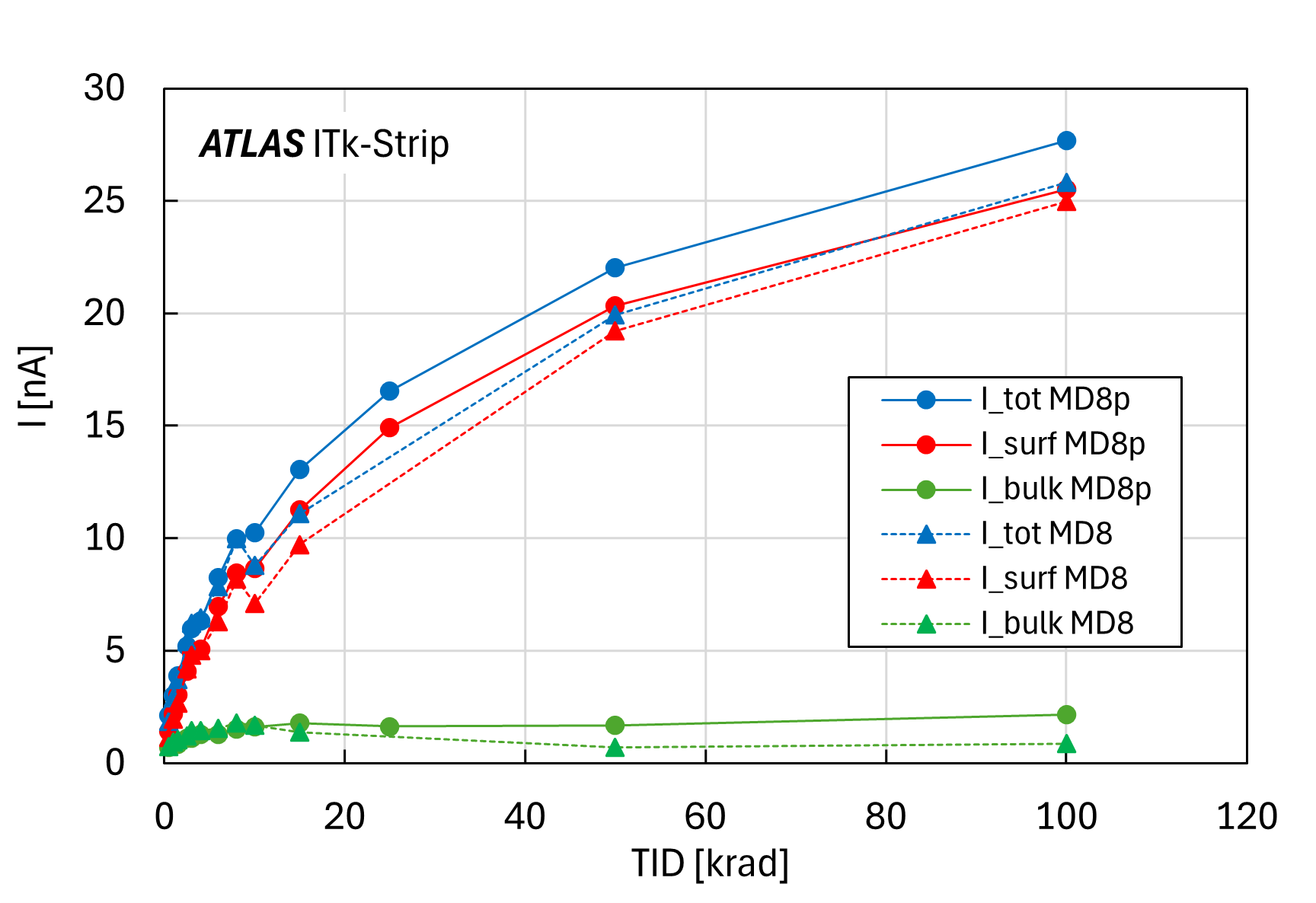}
\caption{(a) Dependence of total, bulk, and surface currents of MD8 and MD8p diodes on bias voltage measured before irradiation. (b) Total, bulk, and surface currents of irradiated MD8 and MD8p diodes measured at $\mathrm{V_{bias}}$ = 300 V after their annealing for 80 minutes at 60°C. All measurements were done at  {20}~$^\circ$C with a relative humidity~{<1}\%.\label{fig:2}}
\end{figure}

Figure~\ref{fig:2}(a) displays total, bulk, and surface currents as a function of bias voltage measured both on MD8 and MD8p unirradiated diodes. Before irradiation, the bulk and surface currents are of a comparable magnitude and contribute approximately equally to the total current. The bulk and surface currents are slightly larger in MD8p diodes. The cause of the difference is under investigation.

The dependence of total, bulk, and surface currents on TID measured at a reverse bias of 300~V for irradiated MD8 and MD8p diodes after their annealing is presented in Figure~\ref{fig:2}(b). After irradiation, the surface current increases significantly even at low TID values and dominates in the contribution of the total current. The bulk current remains relatively stable up to the maximal measured TID of 100 krad.   The small shift in current vs TID curve at 8 krad is caused by uncertainty in the TID measurements done in two different irradiation campaigns, which is related to the very short irradiation times needed to reach required TIDs. 

\subsection{Annealing studies of bulk, surface, and total currents} \label{sec:currents}
\begin{figure}[h]
\centering
\includegraphics[width=7.4cm]{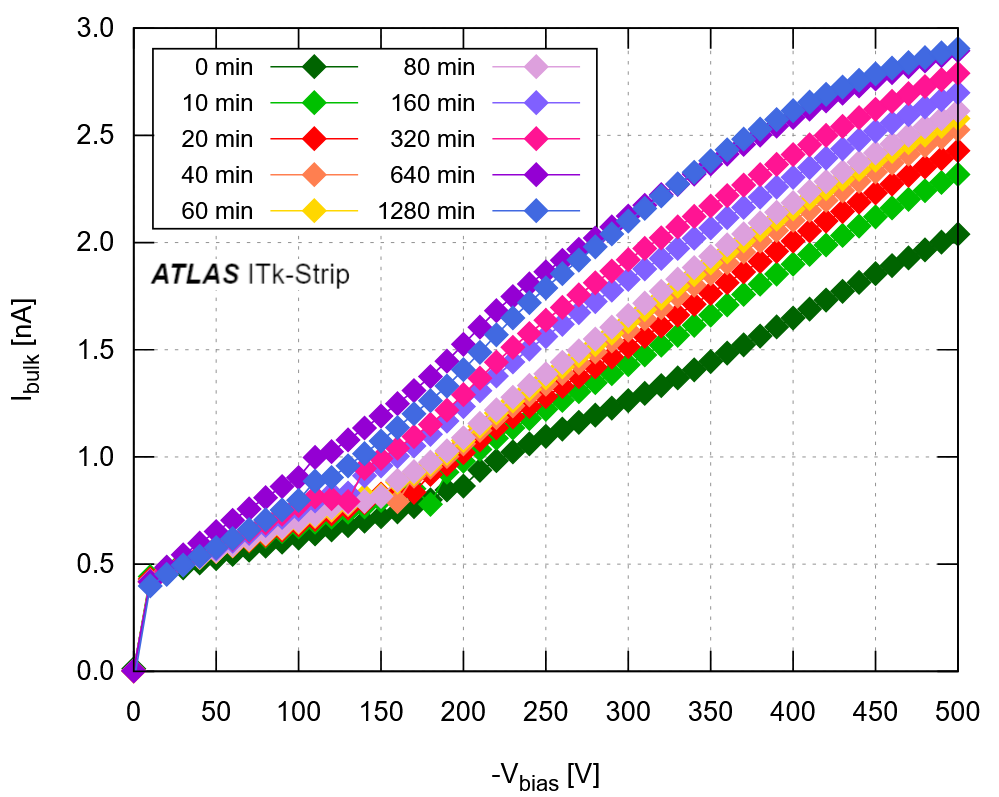}
\includegraphics[width=7.5cm]{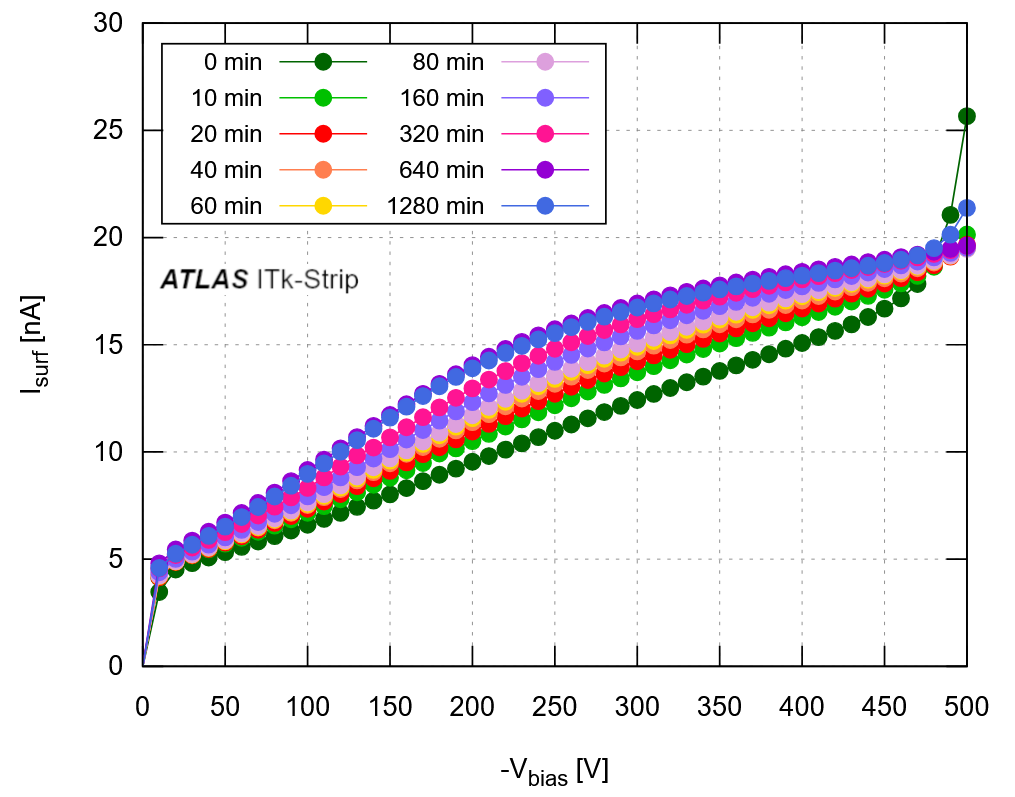}
\caption{Dependence of the bulk (a) and surface currents (b) on the bias voltage of the MD8p diode irradiated to 25 krad measured after the specified times of annealing at  {60}~$^\circ$C. Measurements were performed at  {20}~$^\circ$C with a~relative humidity of \mbox{{<1}\%}.\label{fig:3-annealed}}
\end{figure}

MD8p diode irradiated to a TID of \mbox{25 krad} underwent controlled annealing at  {60}~$^\circ$C for 10, 20, 40, 60, 80, 160, 320, 640, and 1280 minutes. The bulk and surface \textit{I--V} characteristics measured after the specified times of annealing are shown in Figure~\ref{fig:3-annealed}.


\begin{figure}[h]
\centering
\includegraphics[width=7cm]{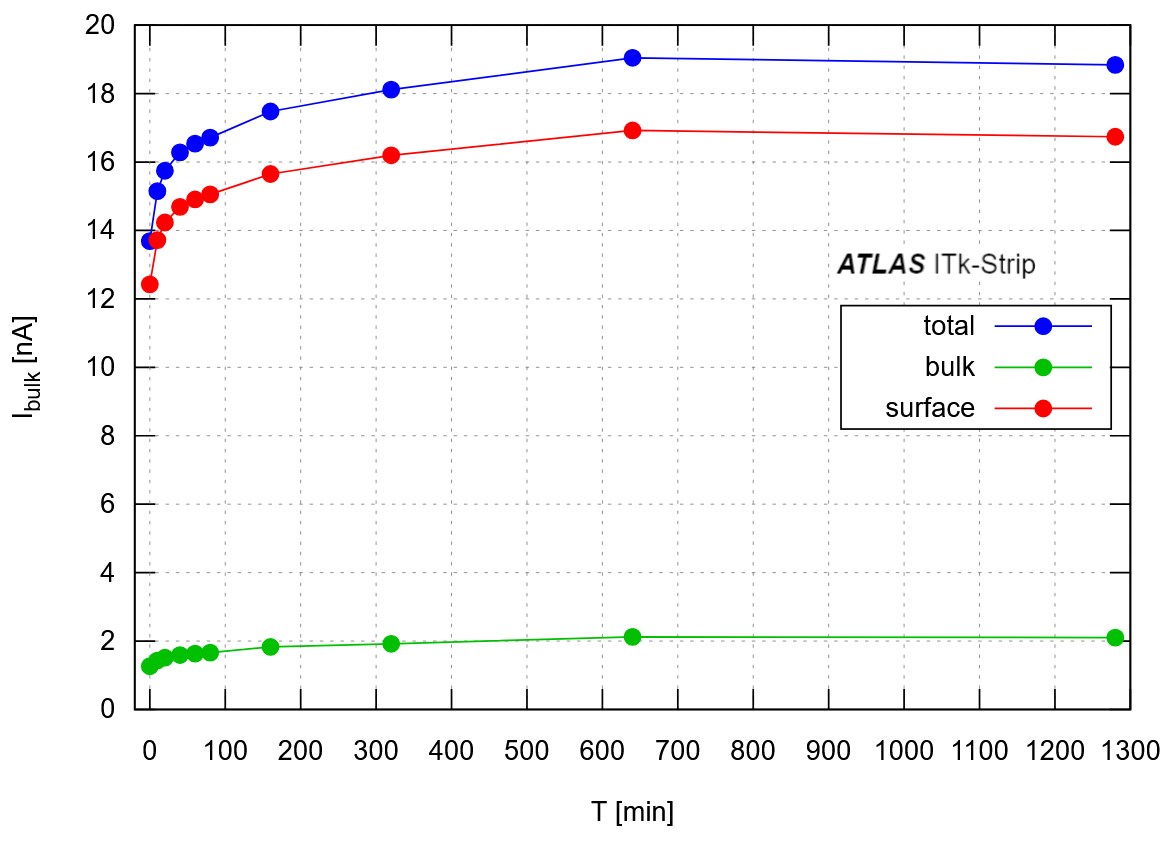}
\includegraphics[width=7cm]{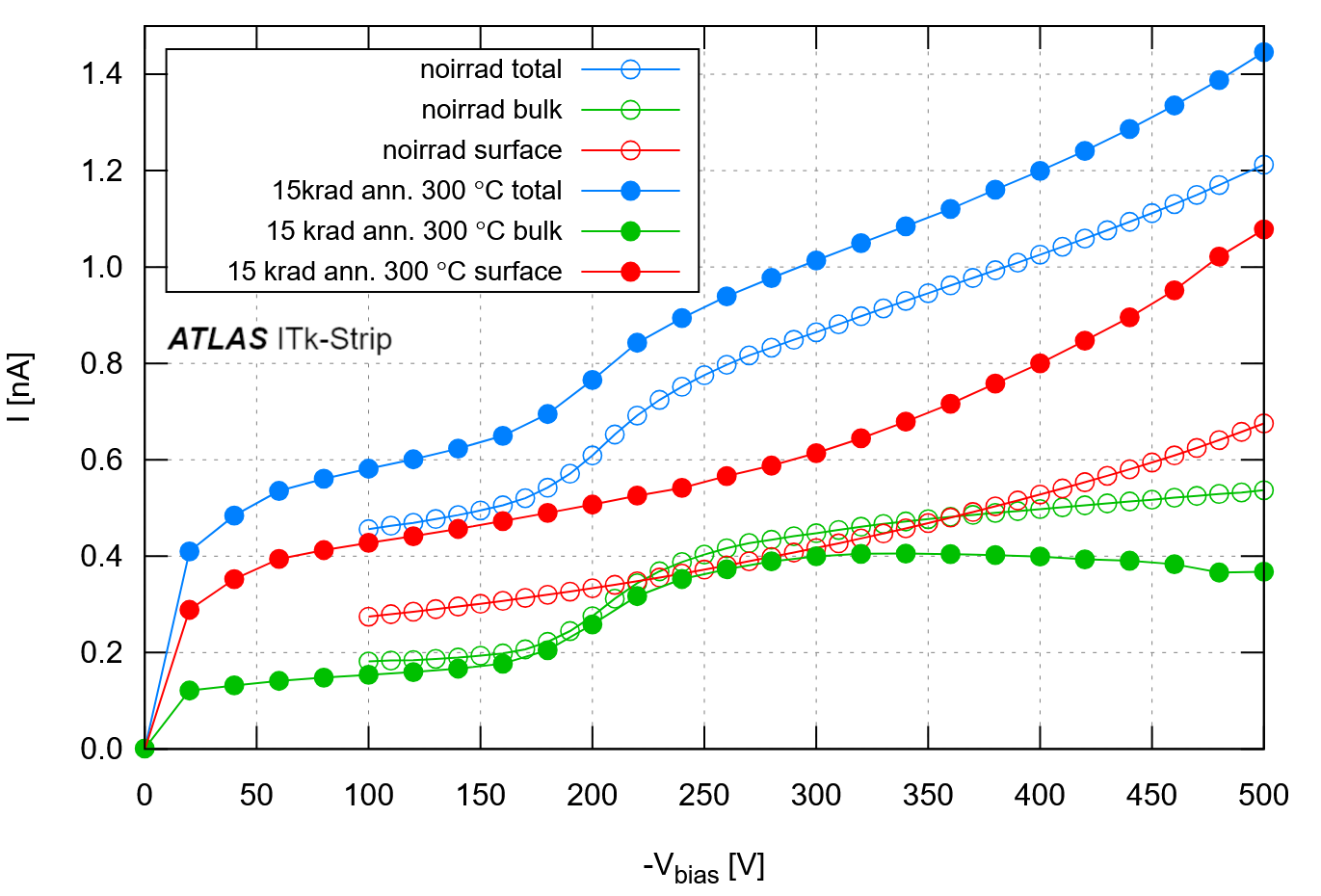}
\caption{(a) The evolution of the total, bulk, and surface currents with the time of annealing at  {60}~$^\circ$C measured for the MP8p diode irradiated to 25 krad and biased to 300 V. (b) Comparison of the total, bulk, and surface currents of not irradiated MD8p diode and MD8p diode irradiated to 15~krad and annealed up to  {300}~$^\circ$C. All measurements were performed at  {20}~$^\circ$C with a relative humidity of \mbox{{<1}\%}.\label{fig:4}}
\end{figure}

The bulk current increases slightly with increasing annealing time at  {60}~$^\circ$C, as shown in Figure~\ref{fig:3-annealed}(a). The surface current also increases with time; however, at higher bias voltages (above $\sim 450\,\mathrm{V})$, the current increase becomes negligible, see Figure \ref{fig:3-annealed}(b). These results suggest that annealing mitigates the soft breakdown of the diode observed before annealing at these bias voltages. The evolution of the total, bulk, and surface currents with annealing time measured at full depletion (at a bias voltage of 300 V) is displayed in Figure \ref{fig:4}(a). The surface and bulk currents seem to saturate at about 640 minutes of annealing at  {60}~$^\circ$C.  

\begin{figure}[h]
\centering
\includegraphics[width=8.0cm]{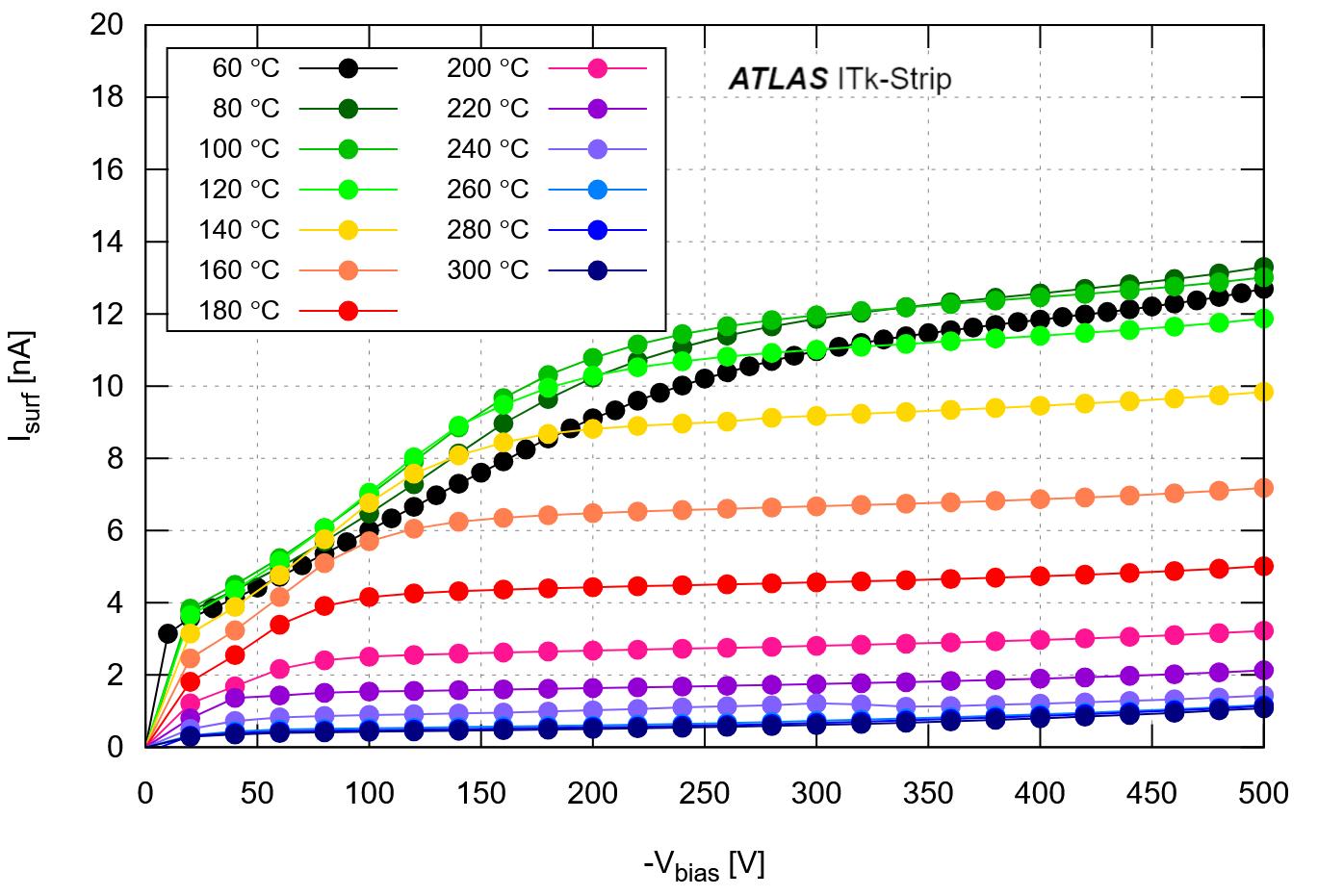}
\includegraphics[width=6.4cm]{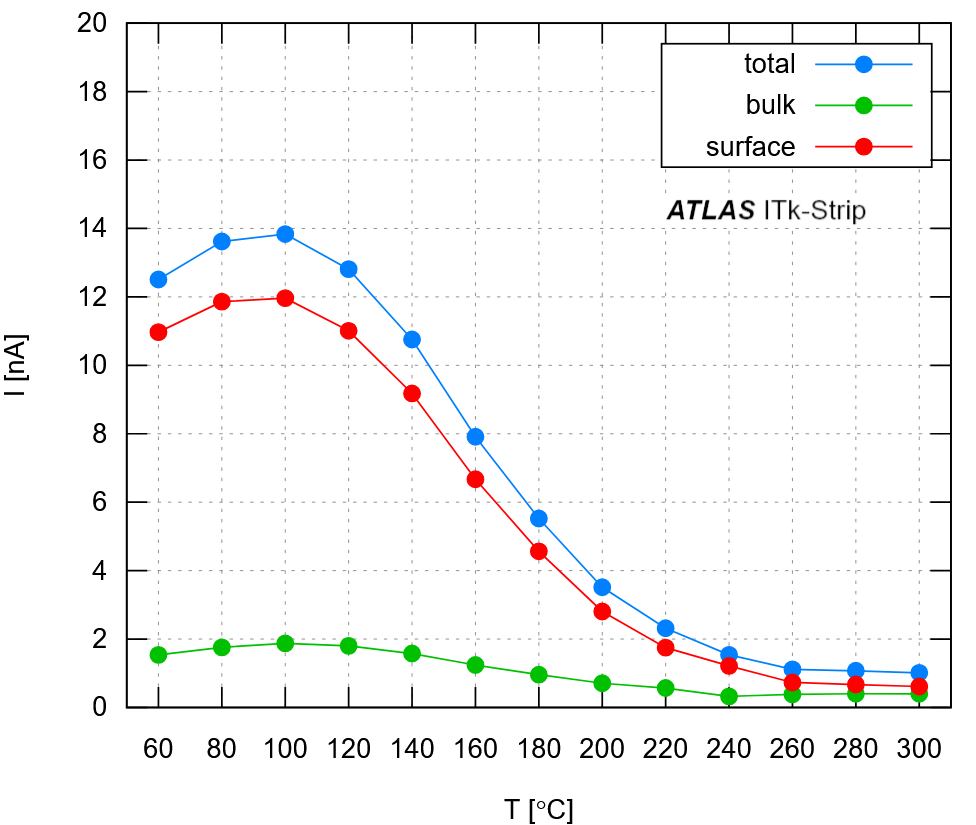}
\caption{(a) The surface currents of the diode MD8p diode irradiated to 15~krad measured after its isochronal annealing with 20-minute steps at temperatures ranging from  {80}~$^\circ$C to  {300}~$^\circ$C. Measurements were performed at  {20}~$^\circ$C with a relative humidity of \mbox{{<1}~\%}. (b) The evolution of total, bulk, and surface currents of MD8p diode irradiated to 15~krad and biased to 300~V with annealing temperature.\label{fig:5}}
\end{figure}

The MD8p diode irradiated to TID of 15~krad initially underwent annealing for 80 minutes at  {60}~$^\circ$C, followed by isochronal annealing with 20-minute steps and temperatures ranging from  {80}~$^\circ$C to  {300}~$^\circ$C. The surface \textit{I--V} characteristics measured after each annealing step are shown in Figure~\ref{fig:5}(a). The annealing behavior of the total, bulk, and surface currents measured at a bias voltage of 300~V is displayed in Figure~\ref{fig:5}(b). As can be seen, the currents exhibit an increase up to  {100}~$^\circ$C. However, at higher annealing temperatures the currents decrease significantly, returning to levels comparable to those before irradiation. This is demonstrated by Figure~\ref{fig:4}(b), which compares the \textit{I--V}s of the total, bulk, and surface currents before irradiation to those measured after irradiation and annealing at  {300}~$^\circ$C. These results show that the radiation-induced defects caused by $\gamma$-irradiation are completely annealed at high temperatures. For comparison, in reference \cite{bib:liao}, where $n^+$-in-$p$ diodes irradiated by $\gamma$-rays to TIDs in range 100 - 200 Mrad were studied, the leakage current remains stable up to  {200}$^\circ$C, then steeply increases till the temperature of   {260}$^\circ$C, and subsequently decreases sharply to the maximum studied temperature of  {300}$^\circ$C.  


\subsection{Temperature dependence of total, bulk, and surface currents}
The temperature dependence of the bulk current measured in proton- and neutron-irradiated silicon samples is summarized in \cite{bib:chilingarov}. In our presented study, the temperature dependence of total, surface, and bulk currents was investigated for $\gamma$-irradiated diodes. Measurements were performed over a~temperature range of  {-50}~$^\circ$C to  {20}~$^\circ$C on three MD8p diodes irradiated to the TIDs of $\gamma$-doses of 10, 50, and 100 krad. The measured currents were fitted using the expression: 

\begin{equation}\label{eq:1}
    I(T) = A T^2 \exp \left(-\frac{E_A}{2kT}\right),
\end{equation}

\noindent where $k$ is the Boltzmann constant ($k$ = {8.617 $\times$ 10$^{-5}$  eV.K$^{-1}$). Parameters $A$ and activation energy $E_A$ were treated as free parameters for total, bulk, and surface currents.

The temperature dependence of the surface current of the MD8p diode irradiated to 100~krad is displayed in Figure~\ref{fig:6}.  The left plot shows the surface currents as a function of bias voltage measured at temperatures ranging from  {-50}~$^\circ$C to  {20}~$^\circ$C, while the right plot demonstrates the fitting of total, bulk, and surface currents measured for bias voltage 150~V at various temperatures by the formula \ref{eq:1}.


\begin{figure}[h]
\centering
\includegraphics[width=6.5cm]{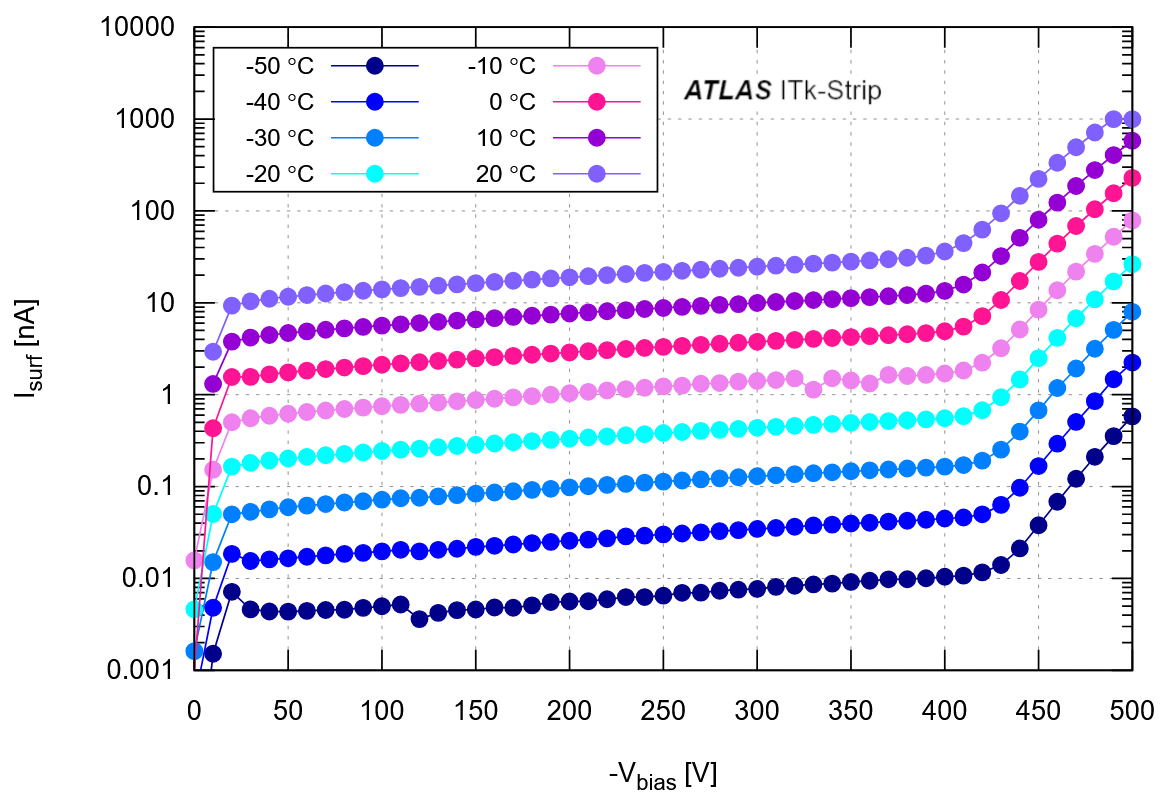}
\includegraphics[width=8.0cm]{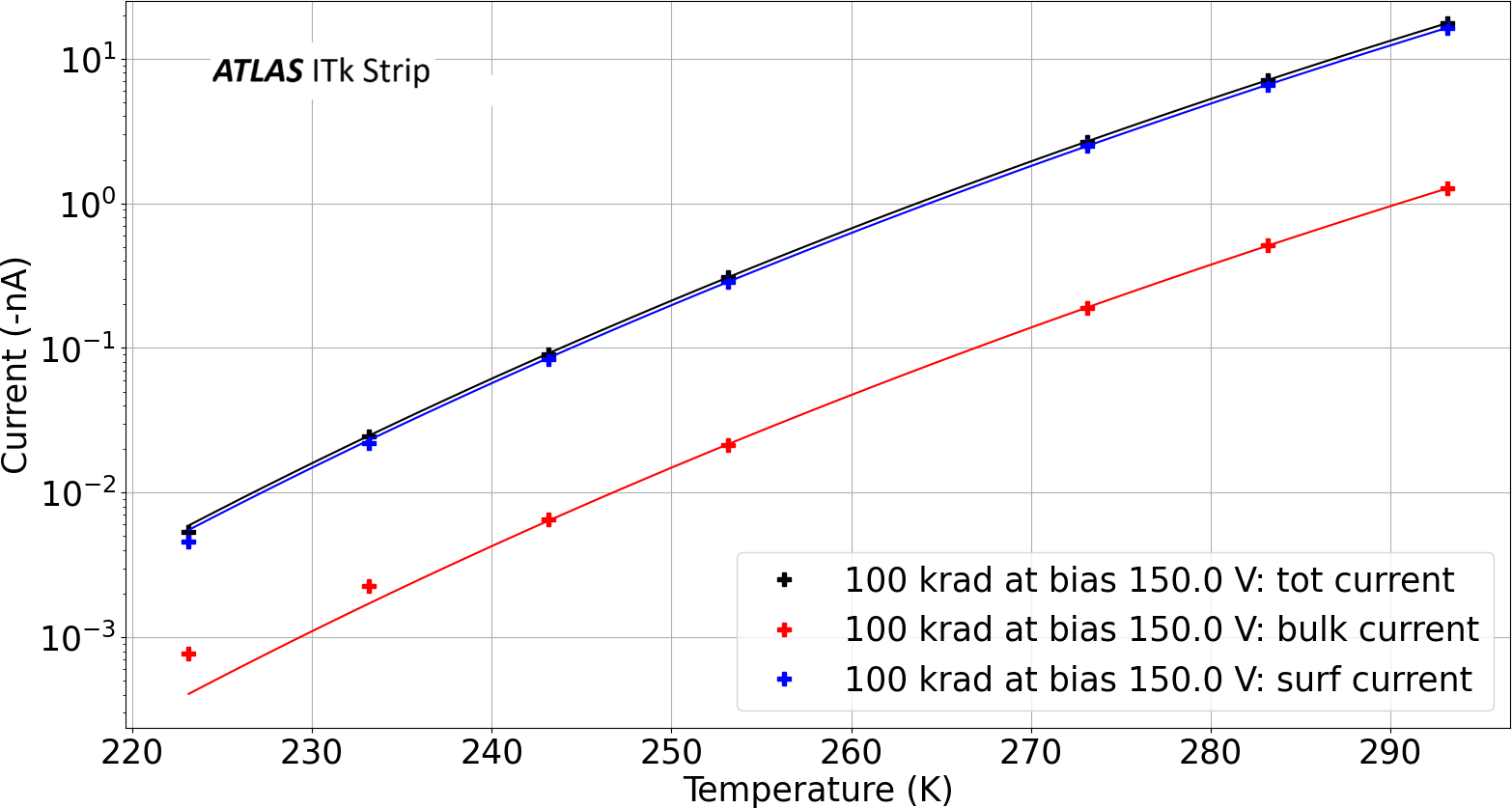}
\caption{(a) Surface currents of the MD8p diode irradiated to 100~krad measured at testing temperatures ranging from  {-50}~$^\circ$C to  {+20}~$^\circ$C, and plotted as a function of bias voltage. (b) Temperature dependence of total, bulk, and surface currents of the MD8p diode irradiated to 100 krad and biased to 150 V. Measured data is fitted by the formula~\ref{eq:1}.\label{fig:6}}
\end{figure}


It was assumed that $E_A$ might vary among $I_\mathrm{BULK}$, $I_\mathrm{SURF}$, and $I_\mathrm{TOT}$ due to a presence of the $\mathrm{SiO_{2}}$ layer on the surface of tested samples. Additionally, $E_A$ could vary with dopant concentration; however, no significant changes in radiation-induced defects in the silicon bulk were expected at the low  $\gamma$-doses used in this study. The results show no differences in the activation energies obtained by fitting the temperature dependence of total, bulk, and surface currents by the formula~\ref{eq:1}. 
Additionally, the activation energy values are consistent across all three samples irradiated with different TIDs, as shown in Table 1, where the fitted values of  $E_A$ and $A$ for all three types of currents - total, bulk, and surface - are summarized for each tested diode.
 The fitted values of the activation energy, averaged over all three samples, are  $E_\mathrm{A(TOT)} = (1.198\pm 0.009)$~eV, $E_\mathrm{A(BULK)} = (1.191 \pm 0.012)$~eV, and $E_\mathrm{A(SURF)} = (1.199\pm0.009)$~eV, which are consistent with the Chiligarov’s result of ($1.209 \pm 0.007)$~eV \cite{bib:chilingarov}.

\begin{table}[h!]
\centering

\begin{tabular}{c|c|c|c|c|c|c}

\toprule

\textbf{TID} & \textbf{$E_{\rm A(TOT)}$} & \textbf{$E_{\rm A(BULK)}$} & \textbf{$E_{\rm A(SURF)}$} & \textbf{$A_{\rm TOT}$} & \textbf{$A_{\rm BULK}$} & \textbf{$A_{\rm SURF}$} \\
{[krad]} & {[eV]} & {[eV]} & {[eV]} & {[nA/K\textsuperscript{2}]} & {[nA/K\textsuperscript{2}]} & {[nA/K\textsuperscript{2}]} \\
\midrule
10  & 1.209 & 1.192 & 1.211 & 2.101 & 0.166 & 1.951 \\
50  & 1.184 & 1.172 & 1.185 & 2.281 & 0.137 & 2.148 \\
100 & 1.201 & 1.208 & 1.200 & 2.285 & 0.360 & 3.930 \\
\bottomrule

\end{tabular}

\caption{Results of the fit: Activation energy $E_A$ and parameter $A$ for for total, surface, and bulk currents.} 
\end{table}


\section{Conclusions}
This study investigates the effects of $^{60}$Co $\gamma$-rays on total, bulk, and surface currents in $n^+$-in-$p$ MD8 diodes from ATLAS18 ITk production wafers, exposed to the TIDs ranging from 0.5 to 100~krad, with the aim of understanding current behavior during the initial operation of the new ATLAS tracker.

Before irradiation, the bulk and surface currents in MD8 diodes are comparable; after \mbox{$\gamma$-ray} exposure, the total current increases mainly due to a rise in surface current, with the bulk current staying constant up to maximum TID of 100 krad. No saturation of radiation-induced surface charges was observed within  the studied range of TIDs, suggesting that the saturation threshold lies between 100 krad and the 66 Mrad value previously reported in Ref. \cite{2}. 


Annealing at 60–100$^\circ$C causes a slight increase in bulk and surface currents, especially at bias voltages close to full depletion. In contrast, annealing above 100$^\circ$C significantly reduces both surface and bulk components, restoring current levels to pre-irradiation values. This suggests that defects induced by \mbox{$\gamma$-irradiation} in both the bulk and surface are effectively annealed at high temperatures.

No differences were observed between the activation energies derived from temperature dependence measured for total, surface, and bulk currents.


\acknowledgments

This work was supported by the European Structural and Investment Funds and the Ministry of Education, Youth and Sports of the Czech Republic via projects LM2023040 CERN-CZ and FORTE - CZ.02.01.01/00/22\_008/0004632, by the US Department of Energy, grant DE-SC0010107, and the Spanish R\&D grant PID2021-126327OB-C22, funded by MICIU/ AEI/10.13039/501100011033 and by ERDF/EU.

Copyright 2025 CERN for the benefit of the ATLAS Collaboration. CC-BY-4.0 license.



\begin{thebibliography}{99}


\bibitem{bib:STRIPTDR}
ATLAS Collaboration,
\emph{TDR for the ATLAS Inner Tracker Strip Detector}, {CERN-LHCC-2017-005; ATLAS-TDR-025. \href{https://cds.cern.ch/record/2257755/files/ATLAS-TDR-025.pdf?version=3}{ATLAS-TDR-25.pdf}}

\bibitem{1}
ATLAS Collaboration,
\emph{The ATLAS Experiment at the CERN Large Hadron Collider},
 JINST \textbf{3} S08003 (2008)
 
 \bibitem{pdg2024}
S. Navas et al. (Particle Data Group), Phys. Rev. D 110, 030001 (2024), pp 621-623; \url{https://doi.rog/10.1103/PhysRevD.110.030001}

\bibitem{Sze}
S. M. Sze, Physics of Semiconductor Devices, 2nd Edition, pp 379-385, JOHN WILEY and SONS, ISBN 0-471-05661-8

 \bibitem{bib:renate}
R. Wunstorf et al.,
\emph{Damage-induced surface effects in silicon detectors},
NIMA377 (1996) 290-297
 

\bibitem{2}
M. Mikestikova et al.,
\emph{The study of gamma-radiation induced displacement damage in  $n^+$-in-$p$ silicon diodes},
NIMA1064 (2024) 169432

\bibitem{3}
I. Zatocilova et al., \emph{Study of bulk damage of high dose gamma irradiated p-type silicon diodes with various resistivities},  JINST \textbf{19} C02039 (2024)

\bibitem{bib:mikestik-vertex}
M. Mikestikova et al., \emph{Gamma irradiation of ATLAS18 ITk strip sensors affected by static charge}, PoS(VERTEX2023) (2024) 026

\bibitem{bib:duden}
E. Duden, \emph{Gamma irradiation of ITk silicon strip modules with early breakdown}, PoS(LHCP2024) (2025) 232

\bibitem{bib:unno}
Y.~Unno et al., \emph{Specifications and pre-production of n$^{+}$-in-p large-format strip sensors fabricated in 6-inch silicon wafers, ATLAS18, for the Inner Tracker of the ATLAS Detector for High-Luminosity Large Hadron Collider},
JINST \textbf{18}, no.03, T03008 (2023),
\url{https://doi.org/10.1088/1748-0221/18/03/T03008}

\bibitem{bib:hamamatsu}
\url{https://www.hamamatsu.com/eu/en.html}

\bibitem{bib:ullan} 
M.~Ullán et al., \emph{Quality Assurance Methodology for the ATLAS Inner Tracker Strip Sensor Production}, NIMA 981 (2020) 164521

\bibitem{bib:ujp}
UJP PRAHA a.s., \url{https://ujp.cz/en/}



\bibitem{bib:esa}
European Space Agency, Total Dose Steady-State Irradiation Test Method, ESCC Basic Specification No. 22900 (2003)

\bibitem{bib:liao}
C. Liao et al., \emph{Investigation of high resistivity p-type FZ silicon diodes after $^{60}$Co $\gamma$-irradiation}, NIMA1061 (2024) 169103

\bibitem{bib:chilingarov}
A. Chilingarov, \emph{Temperature dependence of the current generated in Si bulk}, JINST \textbf{8} P10003 (2013)




\end{thebibliography}
\end{document}